# High-resolution Time-Of-Flight PET with Depth-Of-Interaction becomes feasible: a proof of principle

L Cosentino[1], P Finocchiaro[1,3], A Pappalardo[1,4] and F Garibaldi[2]

[1]) INFN, Laboratori Nazionali del Sud, via S.Sofia 62, 95125 Catania, Italy
[2]) INFN, Sezione di Roma1, Gruppo Collegato Sanità, Viale Regina Elena 299, 00161 Roma, Italy

## 1. Abstract

In this paper we prove that the choice of a suitable treatment of the scintillator surfaces, along with suitable photodetectors electronics and specific algorithms for raw data analysis, allow to achieve an optimal tradeoff between energy, time and DOI resolution, thus strongly supporting the feasibility of a prostate TOF-PET probe, MRI compatible, with the required features and performance. In numbers this means a detector element of 1.5mm x 1.5mm x 10mm, achieving at the same time energy resolution around 11.5%, time-of-flight resolution around 150 ps and DOI resolution even below 1 mm. We stress that such a time resolution allows to increase significantly the Noise Equivalent Counting Rate, and consequently improve the image quality and the lesion detection capability.

These individual values correspond to the best obtained so far by other groups, but we got all of them simultaneously. In our opinion this proof of principle paves the way to the feasibility of a TOF-PET MRI compatible probe with unprecedented features and performance, not only innovative for prostate radiotracer imaging but possibly also for other organs.

## 2. Introduction

Prostate cancer is the most common disease and the second cause of cancer death in men. Precise disease characterization is needed about cancer location, size, extent and aggressiveness. The current standard for diagnosing PCa, namely transrectal biopsy, is performed in an almost blind fashion. A new approach is a multimodality imaging detector that could play a crucial role in diagnosis and follow up, by merging anatomical and functional details from simultaneous Positron Emission Tomography (PET) and Magnetic Resonance Imaging (MRI) scans. MRI is the ideal complement to radionuclide imaging of the prostate because of its soft tissue differentiating power. It can also have adjunct value by using the dynamics of contrast enhancement (Olcott et al. 2009, Garibaldi et al. 2010).

Sub-optimal prostate imaging geometries prevent generic scanners from separating the signal from surrounding organs, as their sensitivity, spatial resolution and contrast are worse than potentially achievable with dedicated prostate imagers. The TOPEM project, funded by Istituto Nazionale di Fisica Nucleare in Italy (INFN), is aimed at developing an endorectal probe capable of performing Positron-Emission-Tomography (PET) in Time-Of-Flight (TOF) mode to operate with new radiotracers, still being compatible with Magnetic-Resonance-Imaging (MRI). Exploiting the TOF capability would considerably improve the Signal-to-Noise-Ratio (SNR) and Noise Equivalent Counting Rate (NECR), and therefore consequently the image quality and the lesion detection capability (Moses 2007, Kadrmas et al. 2009, Karp et al. 2008, Chien-Min 2008). The issue with radio-nuclides imaging techniques for prostate is so far the lack of good specific imaging agents, but new promising radiotracers are showing good performance in animal tests (Mease et al. 2007). We believe that, if successful, the proposed multimodal detector would provide unprecedented capability in diagnosis and follow up of the prostate cancer to detect early stage disease and guide the biopsies, as well as become the main tool to be used in therapy monitoring and follow up.

3) Corresponding author, e-mail: finocchiaro@lns.infn.it
4) now at Microsensor S.r.l., Via Case Nuove 23/A, 95030 Mascalucia (CT), Italy

Such an endorectal probe has to be used in coincidence with an external dedicated detector array and/or a standard PET ring (Olcott *et al.* 2009). The system performance will be dominated by the endorectal detector, with a considerable improvement in both spatial resolution and efficiency with respect to the standard fully external configuration (Olcott *et al.* 2009, Clinthorne *et al.* 2003).

The project is challenging, as it aims at achieving a time precision of a few hundred picoseconds (on collinear coincidences between the two gamma rays originated by the positron annihilation) by exploiting arrays of very compact scintillation detectors of 1.5mm x 1.5mm x 10mm size made from LYSO crystals. In order to achieve these goals the Silicon Photomultiplier (SiPM) represents an almost mandatory photosensor choice (Finocchiaro *et al.* 2008a, Finocchiaro *et al.* 2008b, Finocchiaro *et al.* 2009).

The need of detecting small tumors and correcting for partial volume effects brings up an additional challenge, i.e. achieving the capability of measuring the Depth-Of-Interaction (DOI), namely the impact position inside the detector crystal of the gamma ray along its trajectory, with a few millimeter precision. Several research groups around the world are currently pursuing this goal, as it could revolutionize the impact of the PET imaging performance and effectiveness (Ito *et al.* 2011). Unfortunately TOF and DOI precision are somewhat mutually exclusive, therefore whenever one improves the former, the latter worsens and vice versa, unless one releases the constraint on the granularity (and thus image resolution) (Levin 2002, Vandenbroucke *et al.* 2010, Spanoudaki and Levin 2011, Shibuya *et al.* 2008, Song *et al.* 2010, Maas *et al.* 2009, Yang *et al.* 2009).

In this paper we prove that the choice of a suitable treatment of the scintillator surfaces, along with suitable photodetectors, electronics, and specific algorithms for raw data analysis, allow to achieve an optimal tradeoff between detector element size, energy, time and DOI resolution, thus strongly supporting the feasibility of a TOF-PET probe with the required features and performance. In numbers, we are going to show that a detector array can be built whose individual elements are 1.5mm x 1.5mm x 10mm in size, achieving at the same time energy resolution around 11.5%, time resolution below 150 ps and DOI resolution below 1 mm. We stress that such a time resolution can allow to restrict the field of interest along the collinear trajectories to a few centimeters, thus selecting events from the prostate while reducing the background.

To our knowledge these individual values are close to the best obtained so far by other groups, but we improved all of them simultaneously. In our opinion this proof of principle paves the way to the feasibility of a TOF-PET probe with unprecedented features and performance, not only innovative for prostate radiotracer imaging but possibly also for the breast and brain.

## 3. The probe

The probe we are going to design should be made of 450 (15x30) scintillator detectors, each one consisting of 1.5mm x 1.5mm x 10mm LYSO crystals, with an overall active area of about 25mm x 50mm. (see sketch in figure 1).

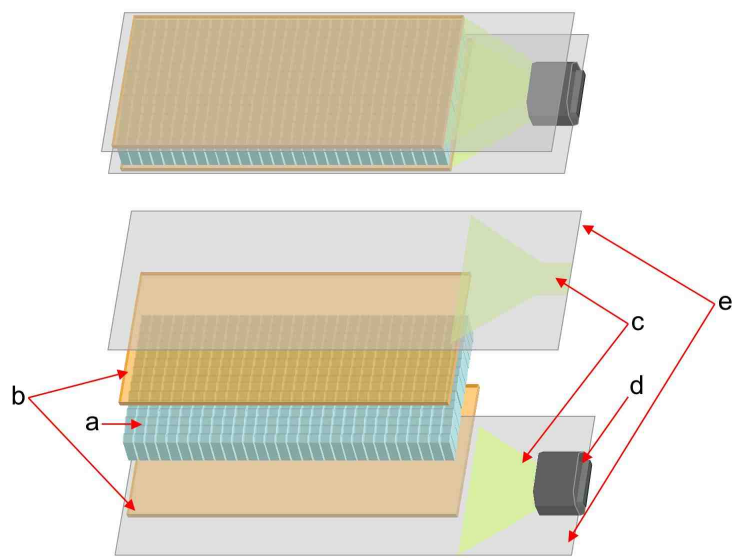

Figure 1. Sketch of the TOF-PET endorectal prostate probe. (a) LYSO scintillators; (b) two arrays of SiPM photodetectors; (c) contacts; (d) connector; (e) backplanes.

The LYSO material was chosen because of its high average atomic number Z, good light yield and rather short light decay time (τ ≈ 40 ns). Each scintillator will have two SiPMs coupled to its ends, in order to collect as much scintillation light as possible thus optimising the energy resolution (see figure 2). The detailed features of the LYSO scintillator can be found in (Qin *et al.* 2005, Chen *et al.* 2007, Loudyi *et al.* 2007). The advantage of using SiPMs, apart from their compactness, low bias voltage (30-70 V) and insensitivity to magnetic fields, is their very fast response (around 1 ns). Together with a fairly good photon detection efficiency, this lets us envisage overall performance of timing and energy resolution comparable to (or even better than) high quality vacuum photomultipliers. The role of each detector consists in detecting the 511 keV gamma rays emitted by the radiotracer accumulated inside the tissue under examination, in coincidence with one of the external detectors. The probe must be able to allow a precision reconstruction of the position of the emitting source, as well as selecting the region of interest to be analyzed by means of the time of flight, thus improving the signal to background ratio (Moses 2007, Kadrmas *et. al* 2009, Karp *et. al* 2008).

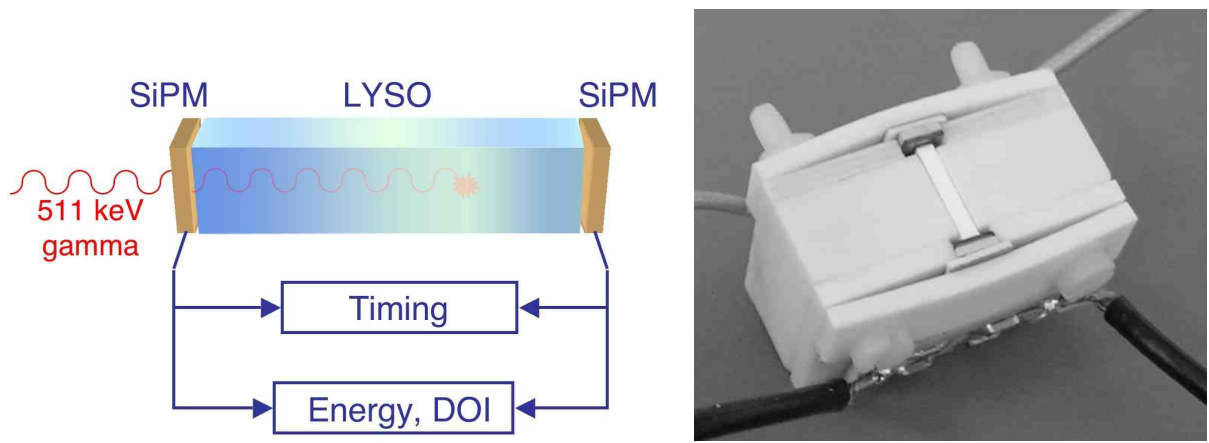

Figure 2. Lefthand side: sketch of the single detector element, made from a LYSO scintillator and two SiPMs; the output signals are suitably handled in order to obtain the Timing, Energy and Depth Of Interaction information. Righthand side: the mechanical assembly of the tested prototype detector element, including the two SIPM polarization networks.

In this work we concentrate on proving that the performance mainly required to the single detection element in terms of energy, time and DOI resolution is feasible and can be attained. We just want to remind here that DOI reconstruction is mandatory if one wants to correct the data for the parallax effect caused both by an extended source and a planar detector array (Ito *et al.* 2011). Several approaches have been pursued by various authors in order to optimize the DOI resolution, none of them so far definitively convincing, either for their cost, complexity and mechanical size, or for the overall performance.

Our approach consists in reading out the scintillation light from the two scintillator ends by means of SiPM photodetectors, and deducing the DOI by comparing the corresponding amounts of charge. Obviously, in order to be sensitive enough, this approach needs to differentiate the amount of transmitted light according to the longitudinal position of production (i.e. the gamma ray impact coordinate). This means that we need to be sensitive to the light attenuation along the crystal, thus implying that the crystal side-faces treatment and the reflector material must be chosen carefully. This way the longer the path, the larger the average number of reflections needed, the smaller the amount of light reaching the photosensor. Unfortunately a considerable attenuation, even though benefitting the DOI sensitivity, spoils the attainable time resolution of a scintillator, as it depends on the photon collection statistics. In other words, one would like to reduce the fraction of transmitted photons per unit length in order to improve the DOI resolution and at the same time one would like to increase the same number in order to improve time resolution. Timing and DOI performances are in mutual competition, as a good time resolution needs that the collected scintillation light be as high as possible, whereas a good DOI resolution requires that the light should be suitably absorbed by reflections on the scintillator surfaces. We have found the tradeoff by adopting a scintillator with polished surfaces and a proper reflector which, along with a suitable electronics and data analysis method, has allowed us to obtain performance that might represent a significant achievement in the technological development of compact TOF-PET devices.

## 4. Experimental Set up

At variance with the setup described in (Garibaldi *et al.* 2010), where we employed polished and teflon-wrapped scintillators coupled with just one SiPM detector, for this work we tested a tiny scintillator (1.5mm x 1.5mm x 10mm, produced by Proteus Inc.) with polished surfaces coated with Lumirror (Huber *et al.* 2001), that gradually attenuates the light propagating inside the crystal by multiple reflections. This condition, as we stated previously, is necessary for the two-sides-weighted DOI measurement.

The two SiPM detectors coupled to the square end surfaces of the crystal (these faces simply polished, no Lumirror) were Hamamatsu S10931-025P(X), with a 3mm x 3mm active area made of 14400 microcells, each one square with 25μm side, and photon detection efficiency (PDE) around 10% in the blue region. The reason why we did not use SiPMs with 50μm elementary cell size (S10931-050P), whose photon detection efficiency is twice as large, is that we did not have any sample available in SMD mount configuration. However, the results obtained with the 25μm-cell sensor can be straightforwardly rescaled to the 50μm-cell one by means of elementary photon statistics considerations.

Our measurements have been performed by employing a $^{137}$Cs gamma source (662 keV), chosen instead of the more appropriate positron-emitting $^{22}$Na which gives rise to pairs of 511 keV gamma rays, just because the activity of the available $^{22}$Na source was too low for a reasonable characterization of our detector. Indeed we made use of 3 cm thick lead collimators of 1 mm diameter, thus strongly reducing the gamma rate on the scintillator. All the resolution values we measured at 662 keV can be easily scaled to the case of 511 keV gamma rays, again by means of the same photon statistics considerations mentioned above.

By means of the collimators we irradiated the scintillator in five different longitudinal positions, namely at 0mm, ±2mm, ±3.5mm referred to the midpoint, and measured the dependence of time and DOI on the position (actually we also performed a set of measurements on three positions using 2 mm diameter collimators, in order to extrapolate the contribution to the DOI resolution due to the collimator size). A sketch of the experimental setup of source, collimator and detector is shown in figure 3, along with a picture of three of the collimators employed.

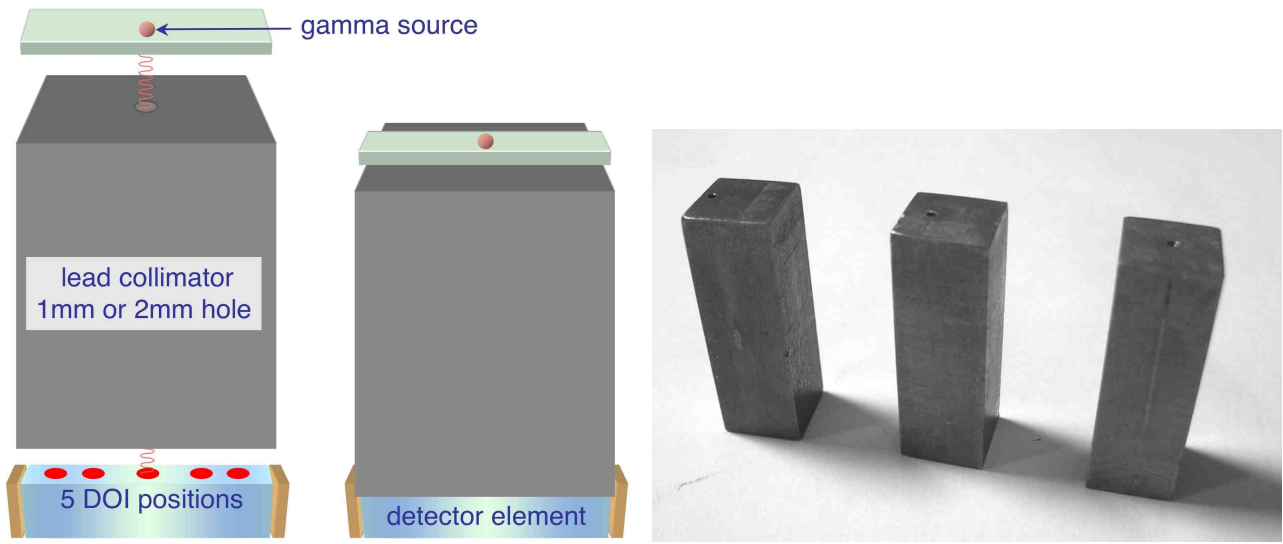

Figure 3. Lefthand side: sketch of the experimental setup of source, collimator and detector.
Righthand side: three lead collimators used for the 5 measurement positions.

The front end electronics included two fast trans-impedance amplifiers (gain 200, 4GHz bandwidth, currently produced by Microsensor S.r.l.), whose outputs were split in two. A copy of these signals was used for charge-to-digital conversion (QDC, for energy and DOI measurements), whereas the other copy was fed into two leading edge discriminators and then used as start-stop signals for a time-to-digital converter (TDC).

In order to perform the measurements we made use of a home-made data acquisition system (DAQ) capable of digitizing and recording simultaneously the two charge signals produced by the SiPM sensors and the time interval between them. The DAQ trigger was set so that the three parameters ($E_{left}$, $E_{right}$, T) were only acquired whenever there was a left-right coincidence within a predefined time window that also created an amount of scintillation light beyond a minimum required threshold, thus enhancing the real gamma events while suppressing uncorrelated noise and spurious signals. A simplified logical scheme of the DAQ system is depicted in figure 4.

The total deposited energy, proportional to the total scintillation light produced, is:

$$E = k\sqrt{Q_L \cdot Q_R} \quad (1)$$

where $Q_L$ and $Q_R$ represent the charge values digitized from the left and right SIPM, and $k$ is a calibration constant. For a more detailed explanation of (1) see Agodi *et al.* 2002.

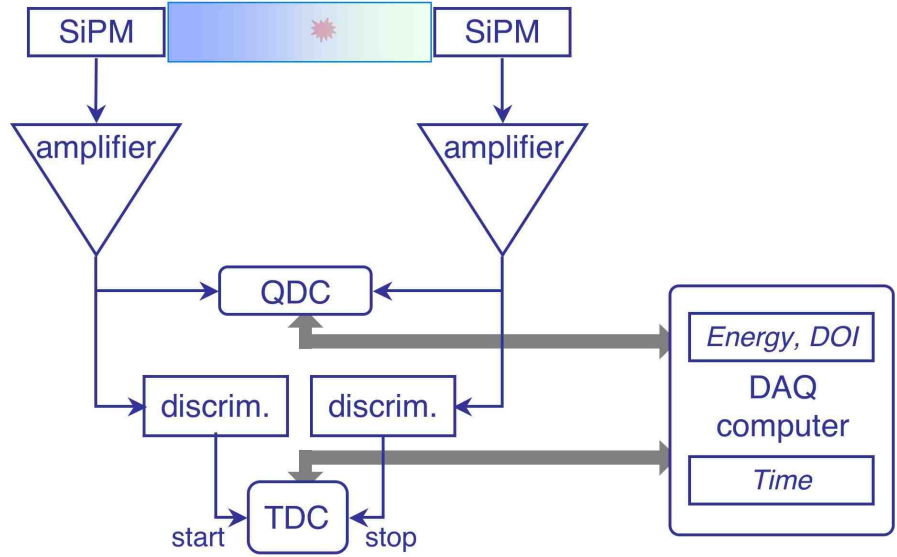

Figure 4. Simplified block scheme of the data acquisition system setup used for the tests.

## 5. Experimental procedure

The left-right time difference was calibrated by correlating the values measured by the TDC when feeding its start and stop with the same pulser signal, being the stop delayed by precisely known time intervals by means of precision cable delay units. All of the measurements were performed with a time calibration of 12.4 ps/channel.

The first step before starting any physical measurement was to choose the supply voltage for the SiPM. After placing the gamma source over the central position, we built the time histogram for several values of the bias voltage ($V_{bias}$), and for each one we measured the FWHM resolution. These values are reported in figure 5, that allowed us to choose the operating bias of 72.9V which minimizes the time resolution. By the way, this was the value recommended by the SiPM manufacturer. We remark that for each measurement the time pickoff on the SiPM signal was chosen at an equivalent threshold of 1.5 photoelectrons (phel), just beyond the massive one-phel dark noise but still very low in order to minimize both the statistical time fluctuations in reaching the threshold value and the time walk of the leading edge discriminators employed.

The parabola-like shape of the plot in figure 5 comes from an initial increase of the PDE with voltage, that gives rise to a better photon statistics thus improving the timing resolution. Then, while still increasing the voltage, the noise becomes predominant and the timing resolution worsens again.

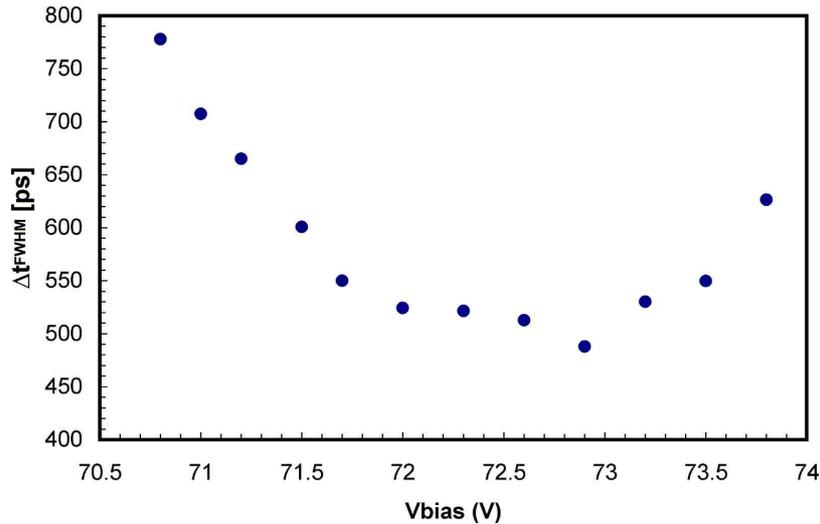

Figure 5. Timing resolution versus SiPM supply voltage. This plot allowed us to choose the operating bias of 72.9V that, by the way, was the one suggested by the manufacturer.

By using (1) we built for each irradiation position the non-calibrated energy histogram. Afterwards we summed these histograms, and the resulting one is shown in figure 6. The peak at 662 keV, clearly visible, has an FWHM (in channels) around 14%. The spectrum is cut at low energy (vertical dotted line), as expected, because of the electronic threshold produced by the aforementioned trigger setting. The Compton shoulder is also visible to the left of the full-energy peak. In the following, whenever we select full-energy events it will mean choosing events whose energy value from (1) falls within an FWHM window around this peak.

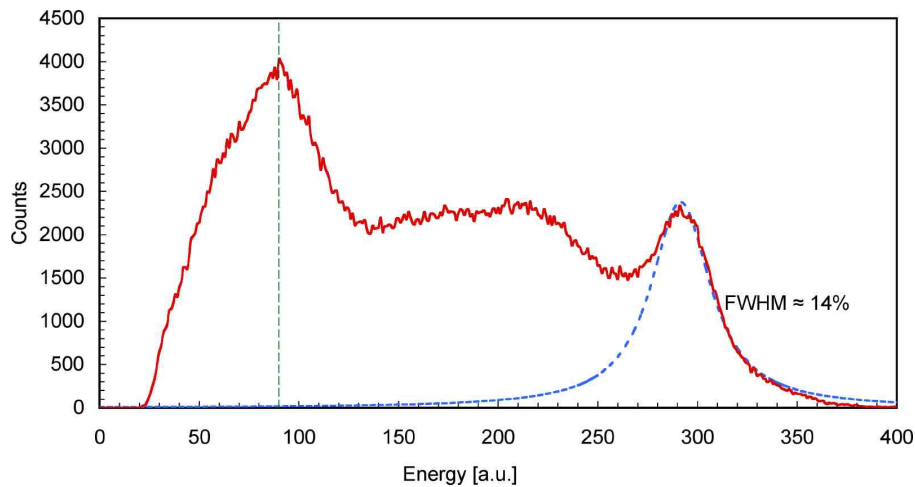

Figure 6. Overall energy spectrum obtained by summing the five energy spectra, corresponding to the five irradiation positions, obtained by means of (1). The resolution around the 662 keV peak (computed in channels) is around 14%. The vertical dashed line roughly indicates the effective electronic threshold, the dotted bell-shaped curve is a lorenzian fit.

For each of the five irradiation positions we built the ($Q_L$ vs $Q_R$) scatter plot, as shown in figure 7 (a-e), where the full energy peaks are clearly visible and move along a well defined geometrical locus defined by $E=E_{peak}$ in (1). figure 8 shows a pictorial 3D representation of the superposition of the scatter plots a, c and e of figure 7 (related to positions 1, 3 and 5), in order to highlight the presence of the full-energy peak and the overall shape of the energy spectrum of figure 6.

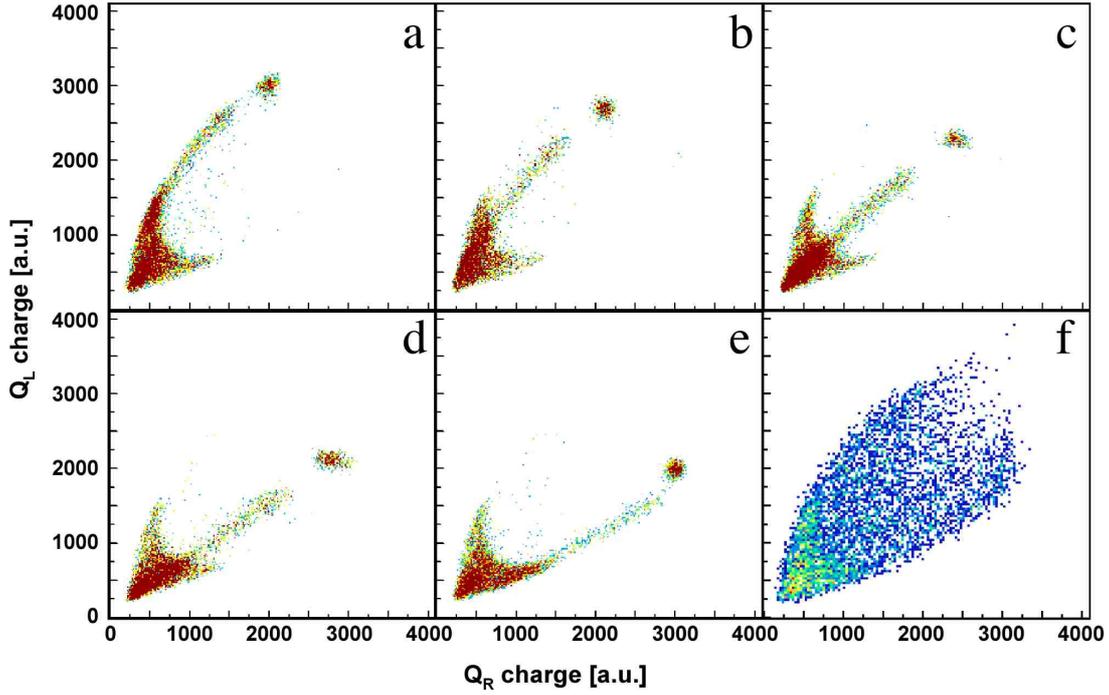

Figure 7. (a, b, c, d, e) Scatter plots ($Q_L$ vs $Q_R$) for the five irradiation positions. The 662 keV full energy peak is clearly visible, and it moves along a locus defined by E=constant in (1). (f) The same scatter plot without gamma source shows the distribution of background counts all over the useful region of the ($Q_L$ vs $Q_R$) plane.

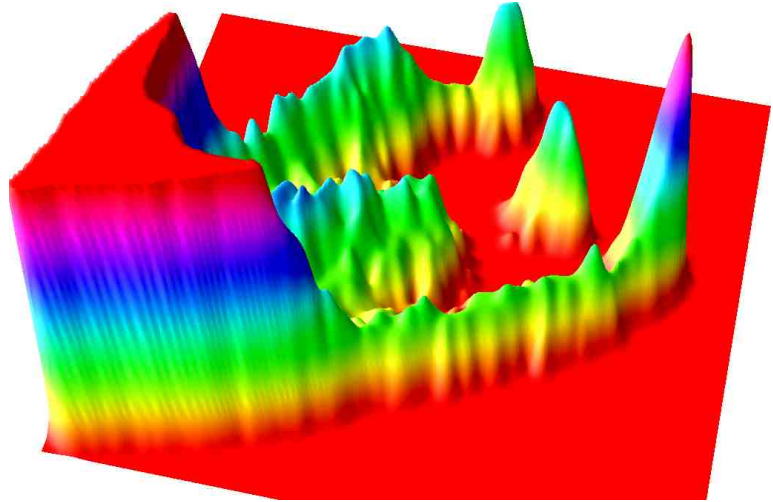

Figure 8. Superposition of the scatter plots (a), (c) and (e) of figure 7, shown in a 3D representation, to highlight the overall shape and the presence of the full-energy peak, according to the spectrum of figure 6.

A check that the experimental apparatus was performing in a clean and symmetrical fashion comes from the evaluation of the attenuation length of the crystal. Since we know the five impact positions of the gamma rays when the full energy is deposited, we can derive the attenuation length ($L_{at}$) using (2):

$$Q(d) = \frac{Q_0}{2} \cdot \varepsilon \cdot PDE \cdot e^{-d/L_{at}} \qquad (2)$$

where $Q(d)$ is the charge measured by the SiPM when the scintillation light of the full-energy peak is produced at a distance $d$, $Q_0$ is the charge corresponding to the total light produced, $\varepsilon$ the optical coupling efficiency, $PDE$ the SiPM photon detection efficiency.

With this method we evaluated two independent values for the attenuation length, using the left and right side SiPM, obtaining quite similar values $L_{atR}$=16.8 mm and $L_{atL}$=16.6 mm (figure 9). This was to be expected, as $L_{at}$ is a property of the crystal with its surfaces and coating, unless there was some hidden problem with the apparatus.

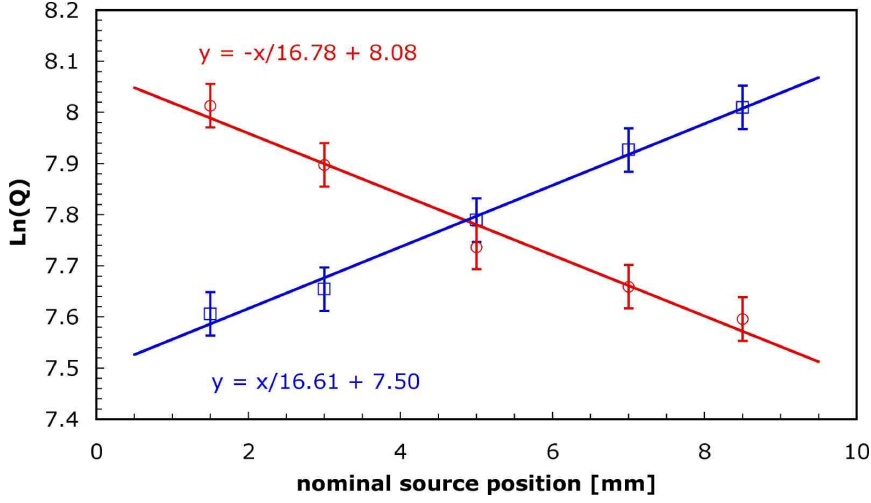

Figure 9. Attenuation plot, built according to (2) using left and right SiPM, that allowed us to deduce two independent values for the attenuation length of the crystal ($L_{atR}$=16.8 mm and $L_{atL}$=16.6 mm), confirming the reliability of the charge measurement with the two SiPMs.

## 6. Data analysis and results

### 6.1. Depth-Of-Interaction

As stated above the goal of this application is to achieve good resolution in DOI and time, therefore we proceed now to explain our DOI performance analysis. The calculation of DOI was done by means of the following formula:

$$DOI_L = M \frac{Q_L}{Q_L + Q_R} \qquad (3)$$

where $M$ is a suitable calibration constant to be determined experimentally by correlating the measured full-energy data with the known irradiation positions via the collimators. The same considerations apply in exactly a symmetrical fashion to the case of $DOI_R$, as the two quantities are strongly bound to each other. From now on we will only use $DOI_L$ for our calculations and assumptions, and will denote it simply $DOI$. Actually this is an approximation, indeed the true formula is:

$$\frac{Q_L}{\sqrt{Q_L \cdot Q_R}} = \sqrt{\frac{Q_L}{Q_R}} = \sqrt{\frac{Q_0 \cdot \varepsilon_L \cdot PDE_L \cdot e^{-d/L_{at}}}{Q_0 \cdot \varepsilon_R \cdot PDE_R \cdot e^{-(L-d)/L_{at}}}} = e^{(L/2-d)/L_{at}} \sqrt{\frac{\varepsilon_L \cdot PDE_L}{\varepsilon_R \cdot PDE_R}} \qquad (4)$$

that, assuming equal left and right values for $\varepsilon$ and $PDE$, allows the calculation of the true DOI (denoted as TDOI) in absolute units

$$TDOI = d = L/2 - L_{at} \cdot ln\sqrt{\frac{Q_L}{Q_R}} \qquad (5)$$

with $L$ the *10 mm* crystal length and $d$ the distance between the impact point and the SiPM under examination. If we used instead (3), and calculated the calibration constant $M$ by using the five known impact positions, we obtained a perfect correlation and thus this result is indistinguishable from the true one of (5). Notice that (3) and (5) provide DOI in the range 0-10 mm, therefore in this coordinate system, that will be used henceforth, the five positions defined by the collimators become respectively 1.5, 3, 5, 7, 8.5 mm.

However, in order for (3) and (5) to hold, $Q_L$ and $Q_R$ must be homogeneous, thus one could decide to equalize the response of the SiPMs as best one can, so that the charge read by the two QDC channels when the source in on the middle position is almost the same. For instance one could bias the SiPMs at the same value and use identical amplifier channels, but small differences in the total gain between the two energy channels will be present anyhow. Even a tiny left-right difference in the optical coupling quality will introduce differences in the output amplitudes, and therefore we decided that the best solution was to equalize $Q_L$ and $Q_R$ in software. Thus event by event we multiplied $Q_R$ by a suitable normalization constant, calculated in order to make the scatter plot of figure 7 symmetric.

By using (5) we can calculate the attenuation length $L_{at}=16.7\ mm$ with a better precision, as shown in figure 10.

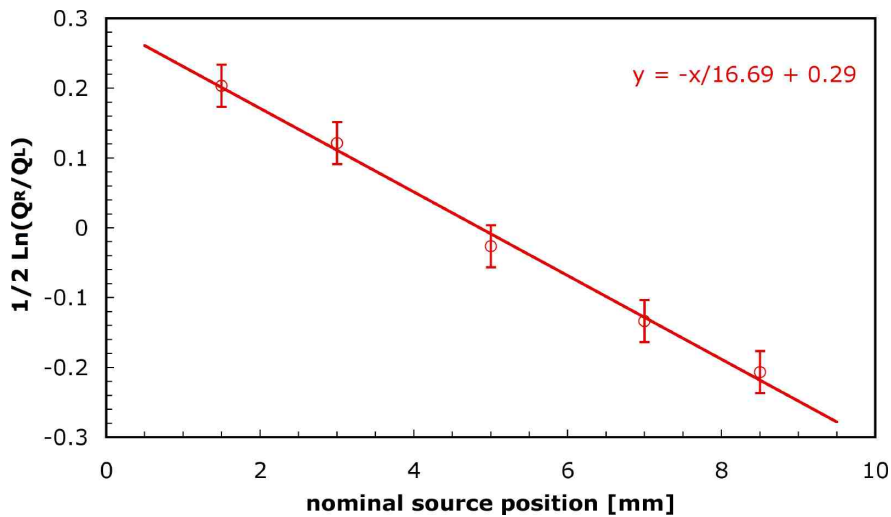

Figure 10. Attenuation plot obtained by means of (5), which allows a better precision as it combines the left and right SiPM data. The attenuation length is $L_{at}=16.7\ mm$.

In figure 11 we show the distribution of the DOI values as a function of the raw position and after calibration in millimeters. It is immediately seen that there was a displacement from the nominal positions, with just the identical behaviour if one uses TDOI (not shown).

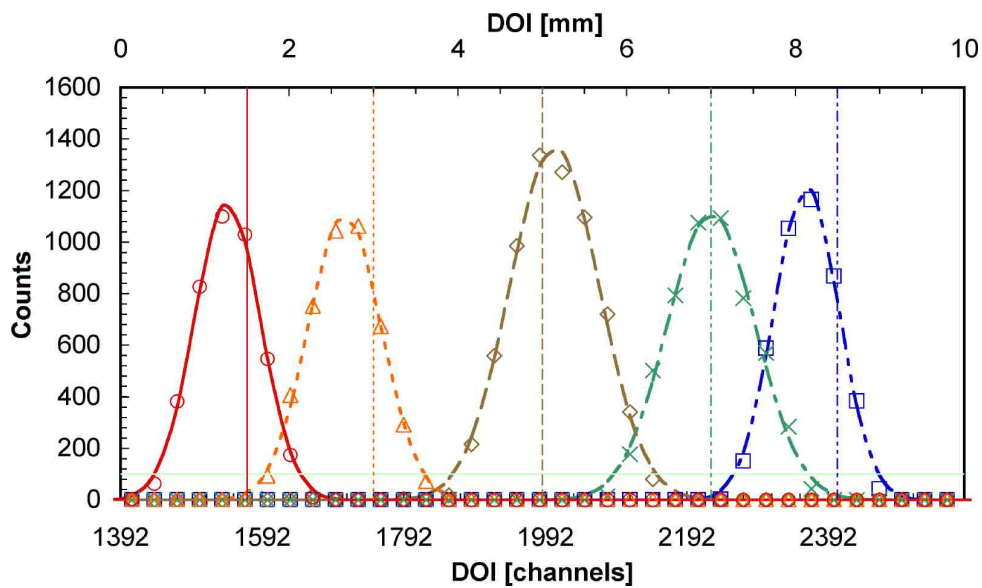

Figure 11. Distribution of the DOI values when separately irradiating the crystal onto each of the five predetermined positions. The symbols represent the data points, the bell-

shaped curves are gaussian fits, the vertical lines are the nominal collimator positions.

In figure 12 we report *DOI* (3) and *TDOI* (5), i.e. the centroids of figure 11 and the corresponding ones for TDOI, as a function of the nominal collimator position. The difference between the two methods is negligible (<0.1 mm), whereas the common overall behaviour is an indication that in some cases there was a real misplacement of the collimator with respect to the crystal and/or of the source with respect to the collimator. This is quite reasonable as one can easily expect in such a tiny setup without a precision alignment system. Since the DOI has to be computed event by event, using (3) is by far simpler and less time consuming than using (5) which implies the calculation of a logarithm and of a square root.

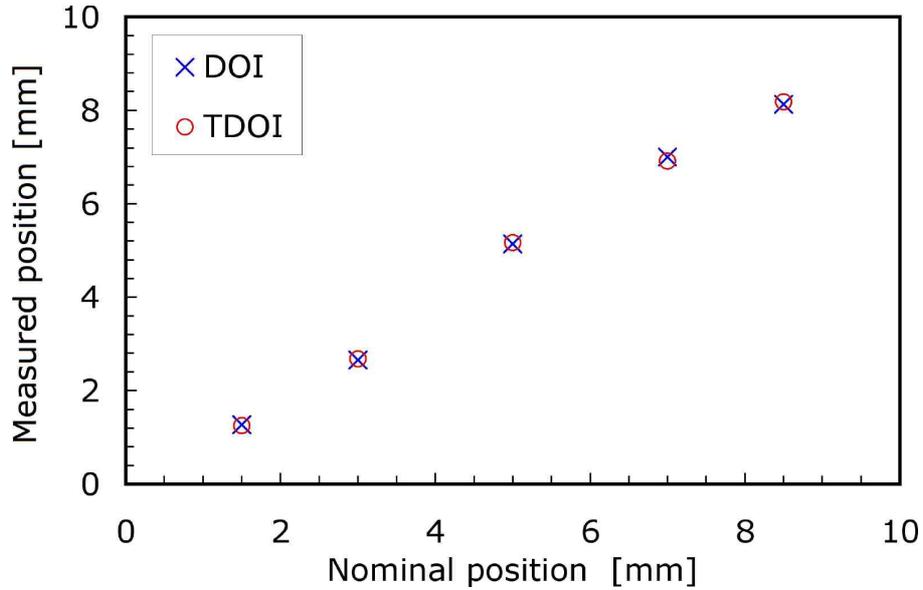

Figure 12. *DOI* and *TDOI*, computed with (3) and (5), as a fuction of the nominal collimator position. The difference between the two methods is negligible (<0.1 mm), whereas the common overall behaviour is an indication of a real small misplacement of the collimator.

Once calibrated the DOI, we were interested in knowing the precision in its measurement. Due to the large number of measurements per position (>3000) the statistical error in the determination of the centroids is negligible, therefore we assumed the measured values as the correct ones and the nominal positions affected by systematic errors. The width of each gaussian in figure 11 is the statistical error to be attributed to each DOI measurement performed on the occurrence of one full-energy gamma detection. With our setup it was reasonable to expect a DOI resolution resulting from the convolution of the intrinsic statistical error of the detector with the collimator aperture, that is >1 mm.

Starting from (3) and using the nominal photon yield (we assume ≈30000 photons/MeV produced), the 16.7 mm average attenuation length, the 10% PDE of the SiPM, one can roughly estimate a typical DOI resolution of the order of $FWHM_{DOI} \approx 1$ mm for impacts near the mid-point of the scintillator. This is quite a result, in our opinion, if compared with the state of the art. Indeed, by using the accurate formula of Vilardi *et al.* 2006, that relates the theoretical TDOI resolution to the above mentioned quantities

$$\sigma_z = \frac{L_{at}}{\sqrt{2N_0}} \sqrt{\left(e^{z/L_{at}} + e^{(L-z)/L_{at}}\right)} \qquad (6)$$

where $z$ is the TDOI coordinate and $N_0$ is the total number of photoelectrons, one obtains the values reported in figure 13 along with the experimental measured ones. We remark that the gamma source we employed is a spherical grain of radioactive material, whose diameter is 1 mm, and since we did not have a precision actuator for positioning it atop the collimator, it is likely that in no case we achieved a perfect collinearity between the source itself and the collimator. The fact that we found values close to 1 mm is an indication that the intrinsic DOI resolution is likely rather smaller than 1 mm.

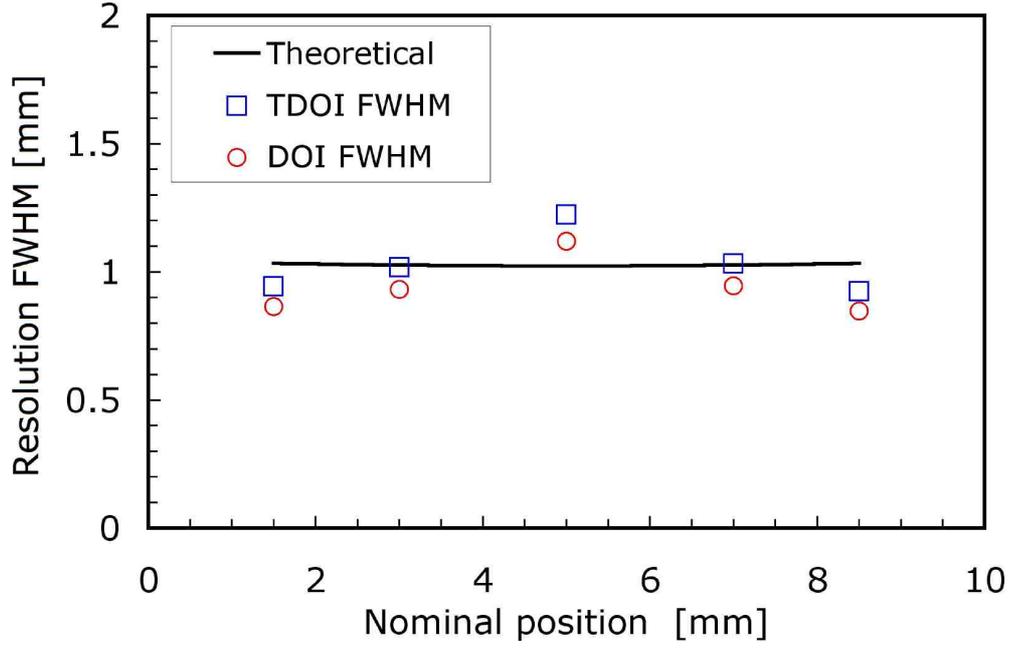

Figure 13. (T)DOI resolution as a function of the nominal collimator position. The theoretical value, about 1.02 mm, is derived from (6).

*6.2. Timing*

In order to study the detector performance in terms of timing we constructed the five histograms representing the distribution of the left-right time difference as detected by the two SiPMs. These plots, built under the condition $|E - E_{peak}| \leq \frac{1}{2} \Delta E_{FWHM}$ and shown in figure 14, are narrow gaussian-shaped distributions ($\Delta t_{FWHM} \approx 500$ ps) whose individual widths are plotted in figure 15. Unfortunately these curves are displaced with respect to each other, the reason being the different propagation time of the light signal inside the crystal itself, as a function of the impact position. This can be clearly seen in figure 16 (a-e), where we show the inclusive scatter plots of DOI versus time for the five selected impact positions (i.e. without any selection on the full-energy peaks). A linear downshift of the time values with increasing impact position is visible.

The plot of figure 16f is the same plot obtained when the source was placed atop a blind collimator, in order to verify the effectiveness of the collimator itself. The contribution of the background counts, as expected, follows the same overall shape but is uniformly distributed with no peaks.

In a real environment without collimators the raw time spectrum will be the projection of the total plot of figure 16 onto the time axis, as shown in figure 17 where we report the sum of the normalized five time distributions corresponding to the different collimator positions. The time resolution in this case is $\Delta t_{FWHM} \approx 910$ ps.

The time shift with DOI position was linearly corrected event by event by making up for the different propagation times. The linear correction coefficients can be easily deduced by fitting the shape of figure 14 or, better, by fitting the position of the centroids of figure 16 as a function of the nominal collimator position. These two plots were rebuilt after applying the just mentioned correction, and can be seen in figure 18 and figure 19. After the correction there is no longer any dependence of time on the impact position, and the overall time resolution, as visible in figure 18, is now about 500ps FWHM.

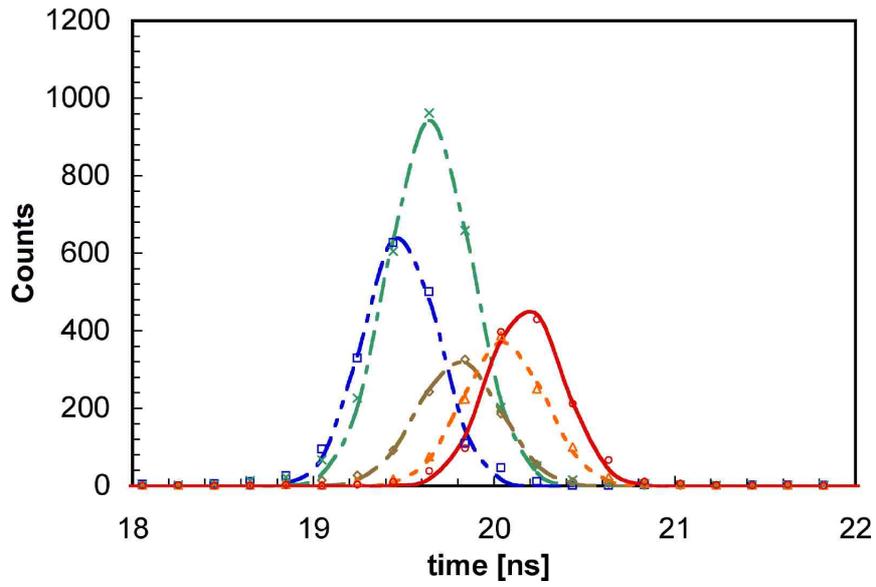

Figure 14. Time spectrum ($T_{left}$-$T_{right}$) for full-energy peak events originated in the five selected positions. The gaussian shaped distributions are narrow but displaced with respect to each other.

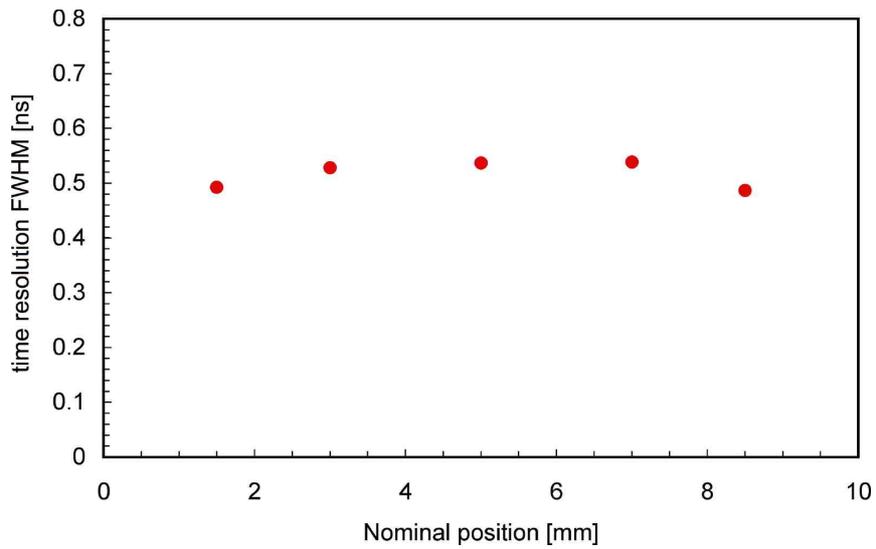

Figure 15. Measured FWHM time resolution (i.e. the widths of the five gaussians of figure 14) as a function of the nominal collimator position.

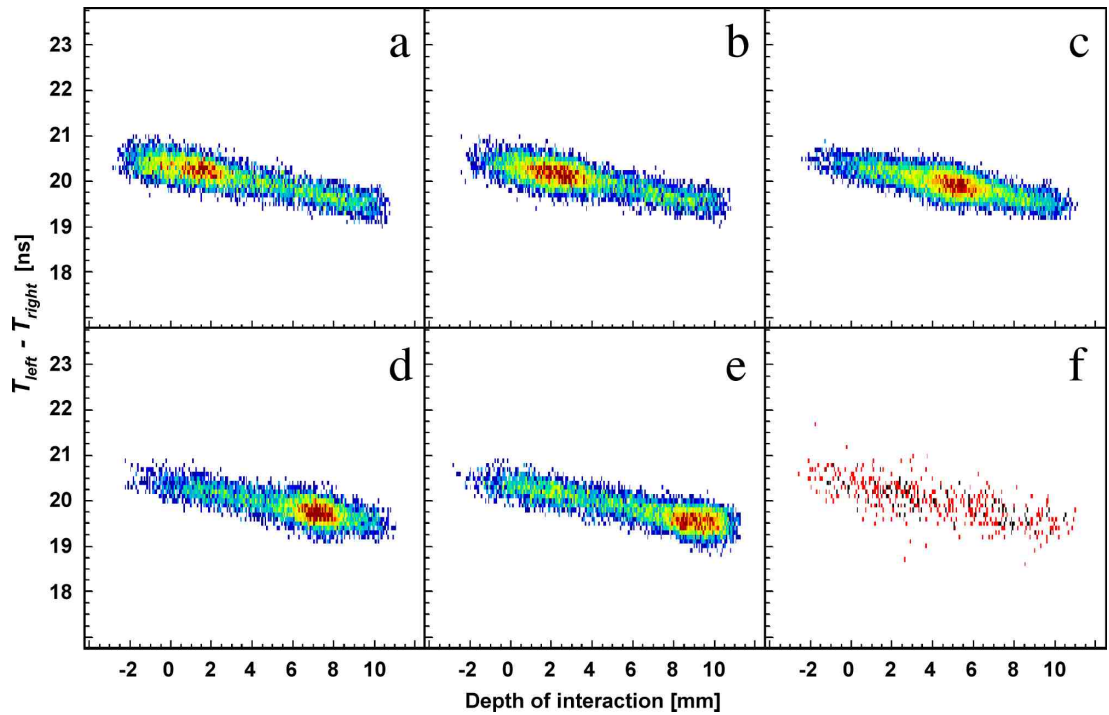

Figure 16. (a, b, c, d, e) Inclusive scatter plot of DOI versus time for the five selected impact positions. The downshift of the time values while the impact position increases is clearly seen. In (f) the same plot when the source was placed atop a blind collimator, in order to highlight the contribution and behaviour of the background counts.

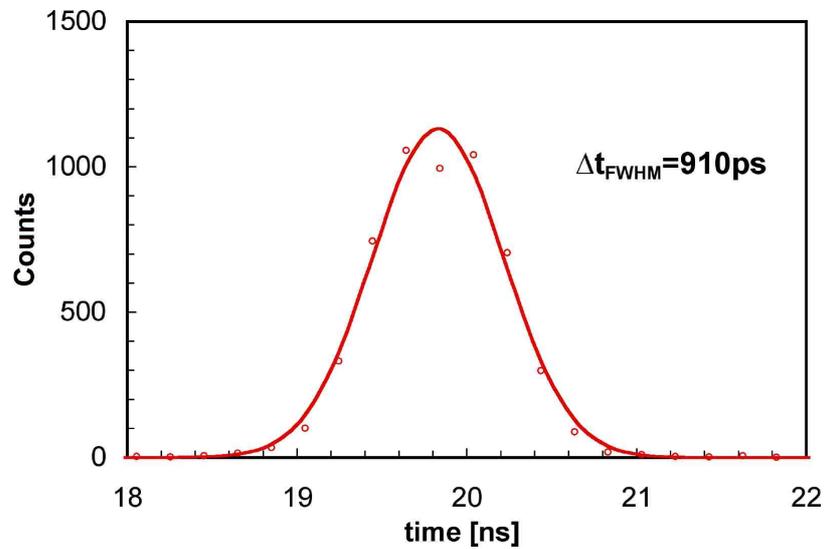

Figure 17. Total raw time spectrum ($T_{left}$-$T_{right}$), as resulting by summing the time spectra for the five different impact positions.

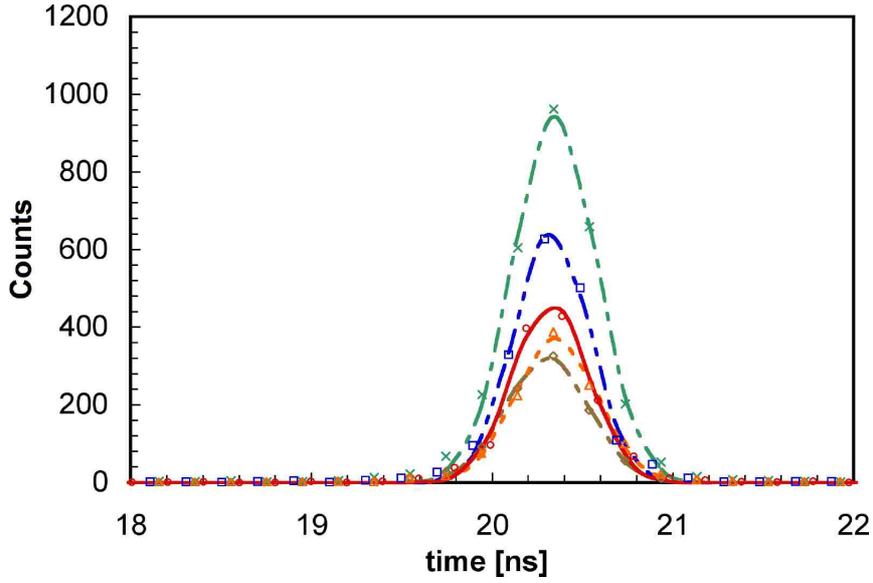

Figure 18. The same plots of figure 14 after correcting for the time shift due to the light-signal finite propagation speed inside the crystal.

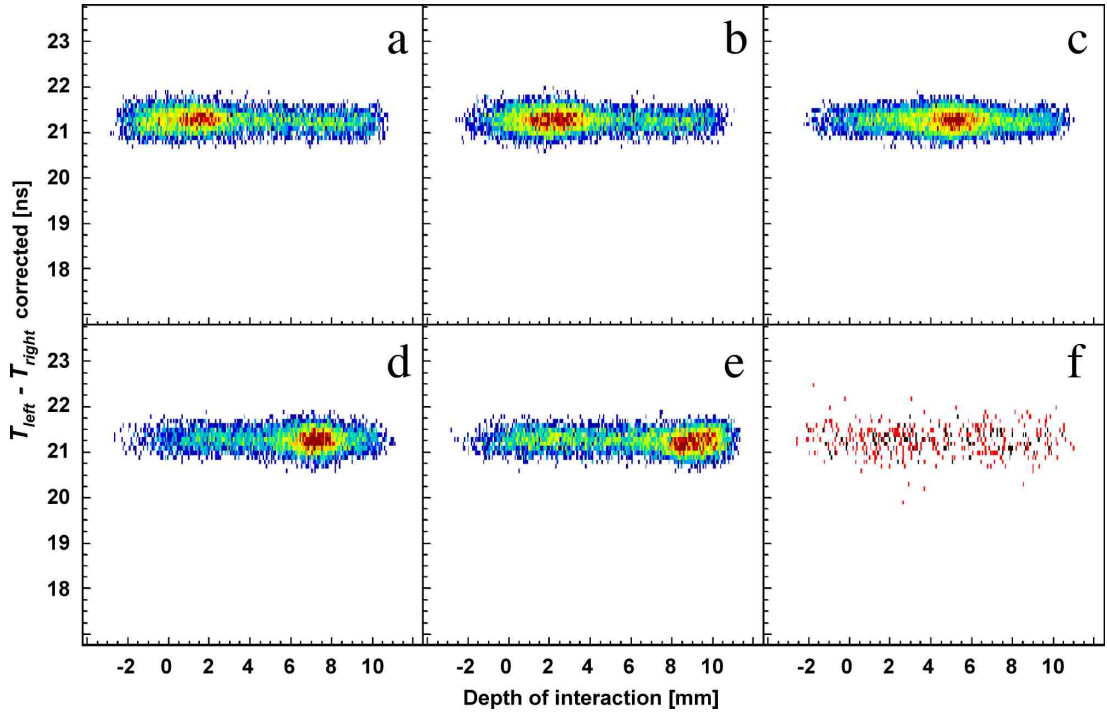

Figure 19. (a, b, c, d, e) The same inclusive scatter plots of figure 16, DOI versus time for the five selected impact positions, after correcting for the time downshift. Also the plot in (f), acquired with a blind collimator, was corrected by the procedure.

By the way, the just explained correction procedure for this time-walk as a function of DOI is basically a measurement of the overall propagation speed of the light signal inside the crystal. Indeed it is also true that

$$x = v \frac{t_R - t_L}{2} \qquad (7)$$

being *x* the position (DOI) and *v* the propagation speed. By reporting these two quantities in the plot of figure 20 and performing a linear fit we were able to measure the propagation speed of the light signal as *v=20.04 mm/ns*.

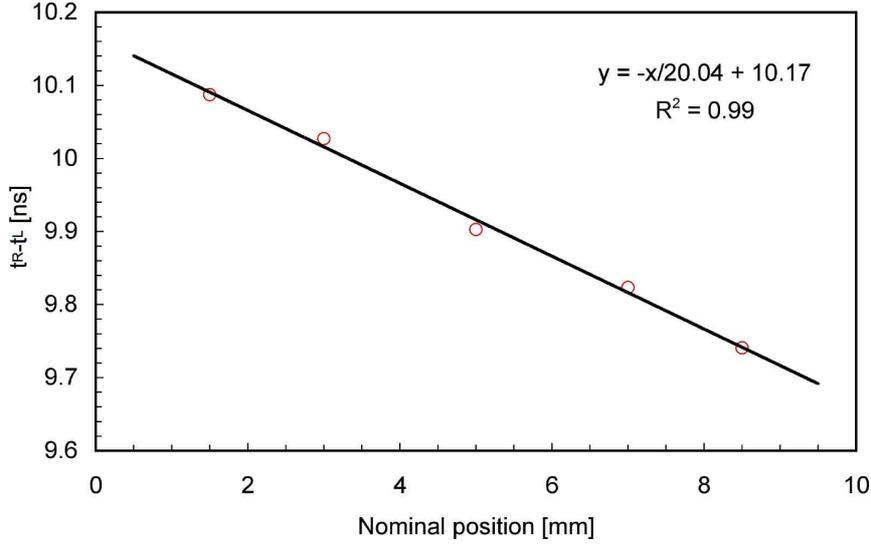

Figure 20. ($t_R$-$t_L$) as a function of the nominal collimator position. From the slope of the linear fit we deduced the propagation speed of the light signal as *v=20.04 mm/ns*.

## 7. Discussion

After characterizing the detector element with a $^{137}$Cs gamma source we now want to scale the measured performances to a real system. This can be done quite effectively, as the scaling depends on well known quantities according to well known statistical laws. In order to scale the detector features to the real case we will assume the gamma energy to be 511 keV, that worsens the overall performance, but we will also assume to employ the 50μm-cell SiPM which doubles the PDE thus improving the performance. The net result in terms of photon statistics is an improvement of a factor $\sqrt{0.511/0.662} \cdot \sqrt{2} = 1.24$.

By making again use of (6) in light of these values one finds out that the expected DOI resolution in a real PET case will scale down to FWHM$_{DOI}$ = 0.82 mm. The energy resolution is expected to improve from 14% to 11.5% FWHM. The issue of time resolution deserves some more detailed comments. Earlier in this paper we quoted our measurement of $\Delta t_{FWHM}$ ≈ 500 ps, and this value comes from the combined contributions of the two SiPMs. Since it is quite reasonable to assume that the two sensors are identical, the contribution of each one is lower by a factor square root of 2, therefore being $\Delta t_{SiPM}$ ≈ 354 ps. If we now scale this value to the expected photon statistics and PDE of the real case we obtain $\Delta t_{SiPM}$ ≈ 285 ps. In the real case, with a real TOF-PET system, for each coincidence event we will have to measure the time difference between two distant detectors of the same kind, say A and B, and this means that we will need to determine two time values $t_A$ and $t_B$. The meaningful time parameter in order to make an effective selection along the line of response (LOR) will thus be:

$$T = \frac{t_B - t_A}{2} \quad (8)$$

because it gives access to the distance *r* along the LOR from the midpoint between the two detectors to where the positron annihilation took place.

$$r = c \frac{t_B - t_A}{2} = cT \quad (9)$$

with *c* being the speed of light.

Since there are two measurements per detector (corrected for the aforementioned time downshift), $t_A$ and $t_B$ will be calculated as averages of the respective pairs of time values:

$$t_{A,B} = \frac{t_{A,B}^{left} + t_{A,B}^{right}}{2} \quad (10)$$

This averaging produces an improvement of the timing precision by square root of 2.

$$\Delta t_A = \Delta t_B = \frac{\Delta t_{A,B}^{left}}{\sqrt{2}} = \frac{\Delta t_{A,B}^{right}}{\sqrt{2}} = \frac{\Delta t_{SiPM}}{\sqrt{2}} = \frac{285\,ps}{\sqrt{2}} \approx 200\,ps \qquad (11)$$

The resolution in $T$ thus becomes

$$\Delta T = \frac{\sqrt{(\Delta t_B)^2 + (\Delta t_A)^2}}{2} = \frac{\Delta t_{A,B}}{\sqrt{2}} = \frac{200\,ps}{\sqrt{2}} \approx 140\,ps \qquad (12)$$

that finally gives rise to a spatial resolution

$$\Delta r = c\Delta T \approx 4.2\,cm \quad FWHM \qquad (13)$$

which allows to select a 4.2 cm segment along the LOR as the origin of the positron annihilation into two gamma rays.

### 8. Operational issues

In this section we are going to make a few operational considerations, basically related to the possible extension of the performance we measured to a real TOF-PET-DOI system with several hundreds detector elements. First of all, we need to make sure that the needed calibrations can be realistically performed without implying complex procedures such as using collimators or other additional mechanical or electronic setups.

The recommended supply voltage will likely be similar but not the same for all the SiPMs but, even in this case, figure 5 clearly shows that there is a comfortable plateau region between 71.5 V and 73.5 V where the timing resolution is rather constant. As for the gain, there will be obviously differences between channels as we stated above, but this will be made up for by suitable normalization constants. These constants can be determined softwarewise in an automatic or semiautomatic fashion every now and then, by means of a gamma source (likely $^{22}$Na) to be suitably placed in a given position with respect to the probe. The same calibration run will be used to measure the attenuation length of each crystal, as the attenuation plot of figure 9 along with the inclusive scatter plot of figure 16, with no collimator, will span the whole available length of the crystals (in our case 10 mm). Once calibrated the TDOI, one will use it to calculate the calibration constant $M$ of (3), in order to use the simpler and faster DOI parameter. The gain and time calibration stability only depend on temperature, and this will be kept stable and under control. The time shift compensation (figure 16) can also be accounted for with the same data acquired in a calibration run with the source. Moreover, in light of the previous considerations we remark that the system could also be calibrated using the same clinical data taken during a scan.

### 9. Conclusion and perspectives.

With this work we have demonstrated that, by adopting suitable tiny detectors, simultaneous high performance in terms of energy, position and, to our knowledge for the first time, time-of-flight and DOI resolution can be achieved in view of a prostate TOF-PET probe application. In the near future we plan to test other kinds of scintillators, like Ca-doped LSO and LSF, in the same configuration, in order to assess which one is the best candidate. By assembling together several hundreds of channels into compact arrays, a prototype probe insensitive to magnetic fields will soon be realized and tested in an MRI-compatible environment. The measured performance of the tested detector element lets us envisage the possibility of a strong increase in the SNR and NECR, thus improving the image quality and the lesion detection capability. In our opinion this represents a proof of principle that paves the way to the feasibility of a TOF-PET probe with unprecedented features and performance, not only innovative for prostate radiotracer imaging but possibly also for other organs.

**Acknowledgments**

We are indebted to several people of the INFN-LNS staff, namely: C. Calì, P. Litrico, S. Marino, G. Passaro for their support with electronics; F. Ferrera for his help with the data acquisition system; B. Trovato, S. Di Modica, M. Tringale, G. Vasta for all the aspects related to micromechanical machining.

We are also deeply grateful to R. Pani for his invaluable suggestions about the collimators and their physical effects on gamma detection.